\documentclass{amia}
\usepackage[labelfont=bf]{caption}
\usepackage{color}

\usepackage[super,comma]{natbib}
\usepackage{times}
\usepackage{soul}
\usepackage{url}
\usepackage[hidelinks]{hyperref}
\usepackage[utf8]{inputenc}
\usepackage{amsmath}
\usepackage{amsthm}
\usepackage{booktabs}
\usepackage{algorithm}
\usepackage{algorithmic}
\usepackage{multirow}
\usepackage{subcaption}
\usepackage{makecell}
\usepackage{amsfonts}
\usepackage{amssymb}
\usepackage{graphicx}

\begin{document}

\setlength{\bibsep}{0.1pt}

\title{STAN-CT: Standardizing CT Image using Generative Adversarial Networks}

\author{
Md Selim$^{1,3}$,
Jie Zhang, PhD$^2$,
Baowei Fei, PhD$^{5, 6}$,
Guo-Qiang Zhang, PhD$^7$ ,
Jin Chen, PhD$^{1,3,4}$}

\institutes{
$^1$Department of Computer Science, University of Kentucky, Lexington, KY\\
$^2$Department of Radiology, University of Kentucky, Lexington, KY \\
$^3$Institute for Biomedical Informatics,  University of Kentucky, Lexington, KY \\
$^4$Department of Internal Medicine,  University of Kentucky, Lexington, KY \\
$^5$Department of Bioengineering, University of Texas at Dallas, Richardson, TX \\
$^6$Department of Radiology, UT Southwestern Medical Center, Dallas, TX\\
$^7$The University of Texas Health Science Center at Houston, Houston, TX \\
}

\maketitle

\noindent{\bf Abstract}
\textit{
Computed tomography (CT) plays an important role in lung malignancy diagnostics and therapy assessment and facilitating precision medicine delivery. 
However, the use of personalized imaging protocols poses a challenge in large-scale cross-center CT image radiomic studies. We present an end-to-end solution called STAN-CT for CT image standardization and normalization, which effectively reduces discrepancies in image features caused by using different imaging protocols or using different CT scanners with the same imaging protocol. STAN-CT consists of two components: 1) a novel Generative Adversarial Networks (GAN) model that is capable of effectively learning the data distribution of a standard imaging protocol with only a few rounds of generator training, and 2) an automatic DICOM reconstruction pipeline with systematic image quality control that ensure the generation of high-quality standard DICOM images. Experimental results indicate that the training efficiency and model performance of STAN-CT have been significantly improved compared to the state-of-the-art CT image standardization and normalization algorithms.
}

\section{Introduction}
Computed Tomography (CT) is one of the widely used modalities for cancer diagnostics~\cite{prince2006medical,mahesh2011fundamentals}. CT provides a flexible image acquisition and reconstruction protocol that allows adjusting kernel function, amount of radiation, slice thickness, etc. to meet clinical requirements~\cite{midya2018influence}. The non-standard protocol setup broadens the scope of CT uses effectively, but at the same time it creates the data discrepancy problem among the acquired images~\cite{ours_aamp}. For example, the same clinical observation with two different CT acquisition protocols may result in images with significantly different radiomic features, esp. intensity and texture~\cite{berenguer2018radiomics,hunter2013high}. As a result, this discrepancy hinders the effectiveness of inter-clinic data sharing and the performance of large-scale radiomics studies~\cite{berenguer2018radiomics}.

The CT image discrepancy problem could be potentially addressed by defining and using the standard image acquisition protocol. However, it is impractical to use the same image acquisition protocol in all the clinical practices, because there are already multiple CT scanner manufactures in the market~\cite{paul2012relationships} and using a fixed protocol for all patients under all situations will greatly limit the use of the CT techniques in diagnosis, staging, therapy selection, and therapy assessment of lung malignancies~\cite{gierada2010effects}. Alternatively, we propose to develop an image standardization and normalization tool to ``translate'' any CT images acquired using non-standard protocols into the standard one while preserving most of the anatomic details~\cite{ours_aamp}. Mathematically, let target image $x$ be an image acquired using a standard protocol, given any non-standard source image $x'$, the image standardization and normalization tool aims to compose a synthetic image $\hat{x}$ from $x'$ such that $\hat{x}$ is significantly more similar to $x$ than to $x'$ regarding radiomic features. 
 
Deep-learning-based algorithms have been developed for image or data synthesis~\cite{huang2018introduction,liang2018ganai}. U-Net is a special kind of fully connected U-shaped neural network for image synthesis~\cite{u-net}. Generative Adversarial Network (GAN) is a class of deep learning models in which two neural networks contest with each other~\cite{huang2018introduction}. Being one of the mostly-used deep learning architectures for image synthesis, GAN has been extended for CT image standardization~\cite{liang2018ganai}. In GANai, a customized GAN model is trained using an alternative training strategy to effectively learn the data distribution, thus achieving significantly better performance than the classic GAN model and the traditional image processing algorithm called Histogram matching~\cite{weeks1999histogram,jain1989fundamentals}. However, GANai focuses on the relatively easier image patch synthesis problem rather than whole DICOM image synthesis problem~\cite{liang2018ganai}.

In the CT image standardization and normalization problem, the synthesized data and the target data must have the common feature space~\cite{liang2018ganai}. This posees two computational challenges of the work: 1) the effective mapping between target images and synthesized images with great pixel-level details, and 2) the texture consistency among the synthesized images. 
In this paper, to address the critical issues in CT image standardization and normalization, we present an end-to-end solution called STAN-CT. In STAN-CT, we introduce two new constrains in GAN loss. Specifically, we adopt a latent space-based loss for the generator to establish a one-to-one mapping from target images to synthesized images. Also, a feature-based loss is adopted for the discriminator to critic the texture features of the standard and the synthesized images. 
%
%
To synthesize CT images in the Digital Imaging and Communications in Medicine (DICOM) format~\cite{mildenberger2002introduction}, STAN-CT introduces a DICOM reconstruction framework that can integrate all the synthesized image patches to generate a DICOM file for clinical use. The framework ensures the quality of the synthesized DICOM by systematically identifying and pruning low-quality image patches. 
In our experiment, by comparing the synthesized images with the ground truth, we demonstrate that STAN-CT significantly outperforms the current state-of-the-art models. In summary, STAN-CT has the following advantages:
\begin{enumerate}
	\item STAN-CT provides an end-to-end solution for CT image standardization and normalization. The outcomes of STAN-CT are DICOM image files that can be directly loaded into clinical systems.
	\item STAN-CT adopts a novel one-to-one mapping loss function on the latent space. It enforces the generator to draw sample distribution from the same distribution where the standard image belongs to.
	\item STAN-CT uses a new feature-based loss to improve the performance of the discriminator.
	\item STAN-CT is effective in model training. It quickly converges within a few rounds of training.
\end{enumerate}

\section{Background}
CT images are one of the key modalities in lung malignancy studies~\cite{rizzo2018radiomics}. The problem of CT image discrepancy due to the common use of non-standard imaging protocols poses a gap between CT imaging and radiomics studies. To fill the gap, clinical image synthesis tools must be developed to ``translate'' CT images acquired using the non-standard protocol into standard images. 
In the domain of deep learning, generator models and GAN models are the main tools for image synthesis.

\subsection*{Image or data synthesis}
Image or data synthesis is an active research area in computer vision, computer graphics, and natural language processing~\cite{huang2018introduction}. By definition, image synthesis is a process of generating synthetic images using limited information~\cite{hall1983testbed}. The given information includes text description, random noise, or any other types of information. With the recent deep learning breakthrough, image synthesis algorithms have been successfully applied in the area of text-to-image generation~\cite{reed2016generative}, detecting lost frame in a video~\cite{kwon2019predicting}, image-to-image transformation~\cite{karras2019style}, and medical imaging~\cite{yu2019retinal}. 

\subsubsection*{U-net} 
U-Net is a special kind of fully connected neural network originally proposed for medical image segmentation~\cite{u-net}. Precise localization and relatively small training data requirements are the major advantages of using U-Net~\cite{u-net}. A U-Net usually has three parts, down-sampling, bottleneck part and up-sampling. The up-sampling and down-sampling parts are symmetric. There are also connections from down-sampling layers to the corresponding up-sampling layers to add lost feature information during down-sampling. However, while an independent U-net is effective for generating the structural details, it suffers from learning and keeping texture details~\cite{ravishankar2017learning}. This issue can be overcome by adopting U-net in a more sophisticated deep generative model called Generative Adversarial Networks (GANs)~\cite{huang2018introduction}.

\subsubsection*{Generative Adversarial Networks}  
Generative Adversarial Networks (GAN) is one of the mostly-used deep learning architectures for data and image synthesis~\cite{huang2018introduction}. A GAN model normally consists of a generator $G$ and a discriminator $D$. The generator (e.g. U-net) is responsible for generating fake data from noise, and the discriminator tries to identify whether its input is drawn from the real data or not. Among all the GANs, cGAN is capable of synthesizing new images based on a prior distribution~\cite{cgan}. However, the image features of the the synthesized data and that of the target data may not fall into the same distribution. The vanilla cGAN may not be directly applied directly to address the CT image standardization problem. GANai is a customized cGAN model, in which the generator and the discriminator are trained alternatively to learn the data distribution, thus achieving significantly better performance than the vanilla cGAN model. However, GANai focuses on the relatively easier image patch synthesis problem rather than whole DICOM image synthesis problem.

\subsubsection*{Disentanglement}
%
In generative models, latent space often plays a vital role in target domain mapping. Appropriate latent space learning is crucial for generating high quality data. 
Disentanglement is an effective metric that provides a deep understanding of a particular layer in a neural network~\cite{zhang2018visual}. Network disentanglement can assist to uncover the important factors that contribute to the data generation process~\cite{higgins2018towards}.

\subsection*{Alternative Training Strategy}
Model training is one of the most crucial parts of GAN because of the special network architecture (i.e. the generator needs to fool the discriminator while the discriminator tries to detect true data distribution from the false one). In the alternative training mechanism, when one component is in training, the other one remains freeze. Also, each component has a fixed number of training iterations. A variant of alternative training was proposed in \cite{liang2018ganai} named fully-optimized alternative training, where the model training is divided into two phases called G-phase and D-phase. In the G-phase, $D$ is fixed, and $G$ needs to achieve a certain accuracy $\theta_G$ or completes the maximum training steps $t_{max}$. In the D-phase, $G$ is fixed, and $D$ needs to achieve a pre-defined performance $\theta_D$ or reaches a maximum training steps $t_{max}$. When one phase is completed, the other phase will begin. The GANai training will continue until an optimal result is achieved or the maximum epochs are reached. Also, instead of performance competing between a single copy of $D$ and $G$, multiple copies of $G$s and $D$s compete with each other. For example, a $G$ needs to fool multiple $D$s before its phase is over. A rollback mechanism is implemented in GANai so that if a component is not able to fool its counterpart within limited steps, it rolls back to the beginning of its phase and starts the training again. This training method has been successfully applied to the CT image standardization problem. In STAN-CT, we will adopt and further advance this training method aiming to achieve better performance. 

\section{Method}
STAN-CT addresses the long-standing CT image standardization and normalization problem. It consists of a modified GAN model with two new loss functions and a dedicated DICOM image synthesis framework to meet the clinical requirements.

\subsection*{Standardizing CT image patches}\label{sec:GAN}
\begin{figure}[bt!]
\centering
\includegraphics[width=.5\columnwidth]{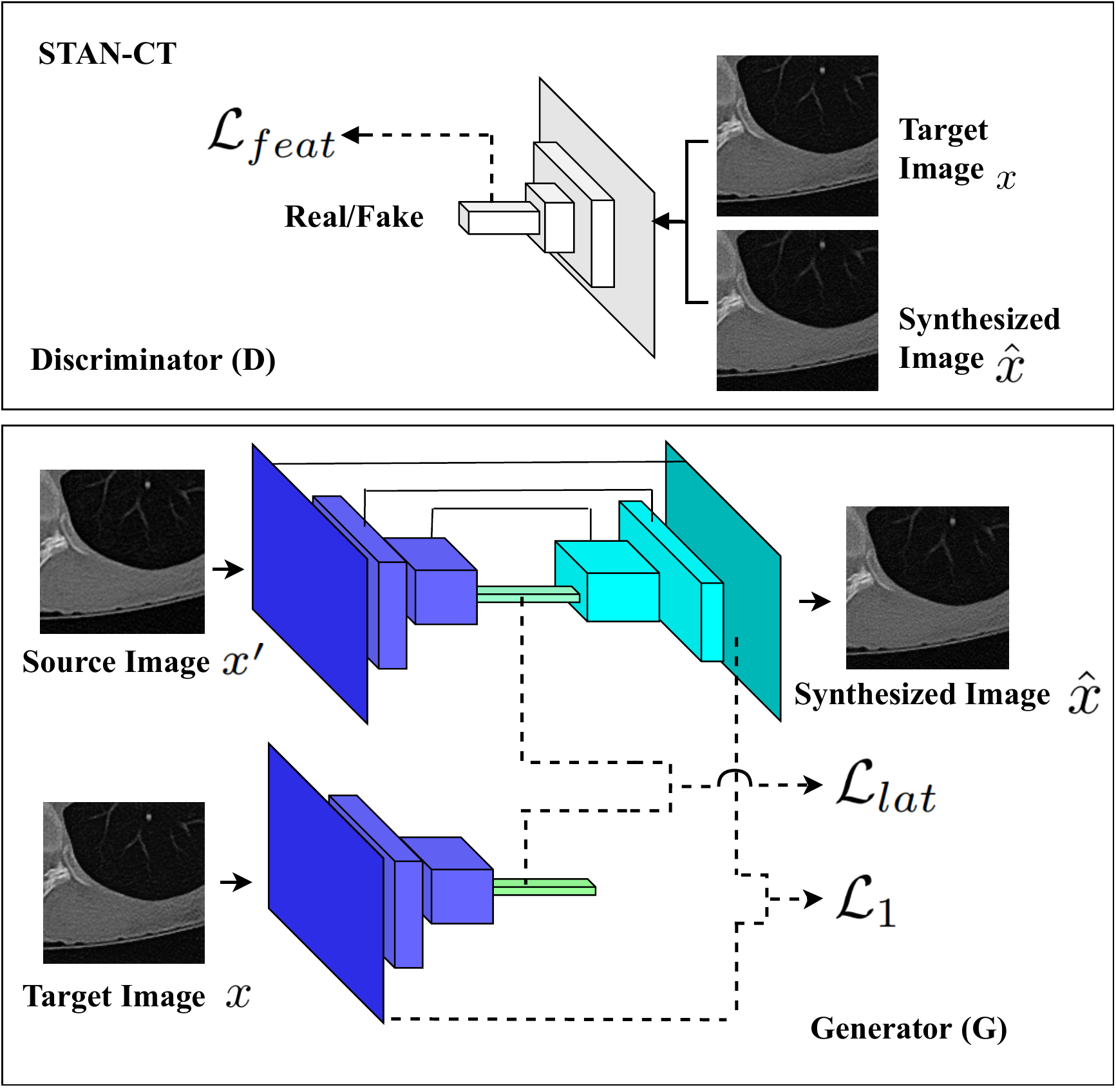}
\caption{ \textbf{GAN architecture of STAN-CT.} The generator $G$ is a U-Net with a new latent loss for synthesizing image patches. The discriminator $D$ is an Fully Convolutional Network classifier for determining whether a synthesized image patch is fake or real.} \label{model}
\end{figure}

Similar to the conventional GAN models, STAN-CT GAN model for standardizing CT image consists of two components, the generator $G$ and the discriminator $D$. $G$ is a \textbf{U}-shaped network~\cite{u-net} consisting of an encoder and a decoder. Both the encoder and the decoder consist of seven hidden layers. There is a skip connection from each layer of the encoder to the corresponding layer of the decoder to address the information loss problem during the down-sampling. $D$ consists of five fully connected convolutional layers. Fig.~\ref{model} illustrates the GAN architecture of STAN-CT.  
Mathematically, let $x$ be a standard image and $x'$ be its corresponding non-standard image. The aim of the generator is to create a new image $\hat{x}$ that has the same data distribution as $x$. Meanwhile, the discriminator determines whether $x$ and $\hat{x}$ are from the standard image distribution.  

\noindent {\bf Loss function of $D$:} In a GAN model, the performance of $D$ and $G$ increases accordingly. We propose to adopt two losses for the discriminator training, i.e. the WGAN~\cite{arjovsky2017wasserstein} adversarial loss function to critic the standard and non-standard images and the fetcher-based loss. 

\noindent {\bf Adversarial Loss.} WGAN is a stable GAN training mechanism that provides a learning balance between $G$ and $D$~\cite{gulrajani2017improved}. STAN-CT adopts the WGAN-based adversarial loss of the discriminator defined as: 

\begin{equation} \label{eq:wgan_d}
\mathcal{L}_{adv(D)} = \triangledown_w \frac{1}{m}\sum_{i=1}^{m}[f(x^{(i)})-f(G(x'^{(i)}))]  
\end{equation}
\noindent where $w$ is the hyper-parameters of $D$, $m$ is the batch size, $x'^{(i)}$ is the input (non-standard image), and ${x^{(i)}}$ is the corresponding standard image.

\noindent {\bf Feature-based Loss.} In addition to the WGAN-based  adversarial loss, STAN-CT introduces a new feature-based loss function $\mathcal{L}_{feat}$. A similar feature-based loss function has been used in~\cite{yang2018low} to improve the generator diversity. Here, we use the feature space of $D$ instead of a secondary pre-trained network to maintain a balanced network (i.e. $D$ and $G$ are not too strong or too weak compared with other). The feature-based loss is described in Eq~\ref{eq:f_loss_d}: 
\begin{equation} \label{eq:f_loss_d}
   \mathcal{L}_{feat} = \mathbb{E}_{(x)}[\frac{1}{V} \left \| \phi (D(G(x')))- \phi (D(x)) \right \|]
\end{equation}
\noindent where $\phi$ is the feature extractor and $V$ is the volume of the feature space, and $G(x')$ is an image generated by $G$ and $x$ is the target image.

Finally, the total loss of D consists of the WGAN-based loss $\mathcal{L}_{adv(D)}$ and the feature-based loss $\mathcal{L}_{feat}$. So, the loss function of $D$ is defined as:
\begin{equation}
\mathbb{L}(D) = \mathcal{L}_{adv(D)} + \lambda_1 \mathcal{L}_{D_{feat}}
\end{equation}
\noindent where $\lambda_1$ is a wight factor ($\lambda_1 \in [0,1]$).

\noindent {\bf Loss function of $G$:} The generator loss consists of three components, i.e. the WGAN-based loss, the latent loss, and the L1 regularization.    

\noindent {\bf WGAN-based loss.} The WGAN-based loss is used to improve network convergence. It is defined as: 
\begin{equation}
\mathcal{L}_{adv(G)} = \triangledown _\theta \frac{1}{m}\sum_{i=1}^{m}f{(G({x'}^{(i)})})    
\end{equation}
\noindent where $\theta$ represents all the hyper-parameters of $G$, $x'{^{(i)}}$ is a source image, and $f$ is 1-Lipschitz function, which returns the Earth-Mover (EM) distance from one metric space to another. 

\noindent {\bf Latent loss.} In the CT image standardization problem, the anatomical properties should be preserved in generating synthesized images. Inspired by~\cite{mao2019mode}, we propose a new latent-vector-based loss function to enforce one-to-one mapping between the synthesized image and the standard image. Specifically, the latent loss $\mathcal{L}_{lat}$ aims to minimize the distance between the latent distribution of the synthesized images and their corresponding standard images.
\begin{equation}
\mathcal{L}_{lat} = \left \| z_{x}-z_{G(x')} \right \| 
\end{equation}
\noindent where $z$ stands for the latent vector, $G(x')$, and $x$ is its corresponding standard image. 

Finally, the total loss of $G$ $\mathbb{L}(G)$ is defined as:
\begin{equation}
\mathbb{L}(G) = \mathcal{L}_{adv(G)} + \lambda_2 \mathcal{L}_{G_{lat}} + \lambda_3 \frac{1}{m}\sum_{i=1}^{m}|x - G(x')|
\end{equation}
where $\lambda_2 \epsilon [0,1]$ and $\lambda_3 \epsilon [0,1]$ are wight factors. $\frac{1}{m}\sum_{i=1}^{m}|x - G(x')|$ is the $\mathcal{L}_1$ regularization function.

\begin{figure}
\includegraphics[width=\textwidth]{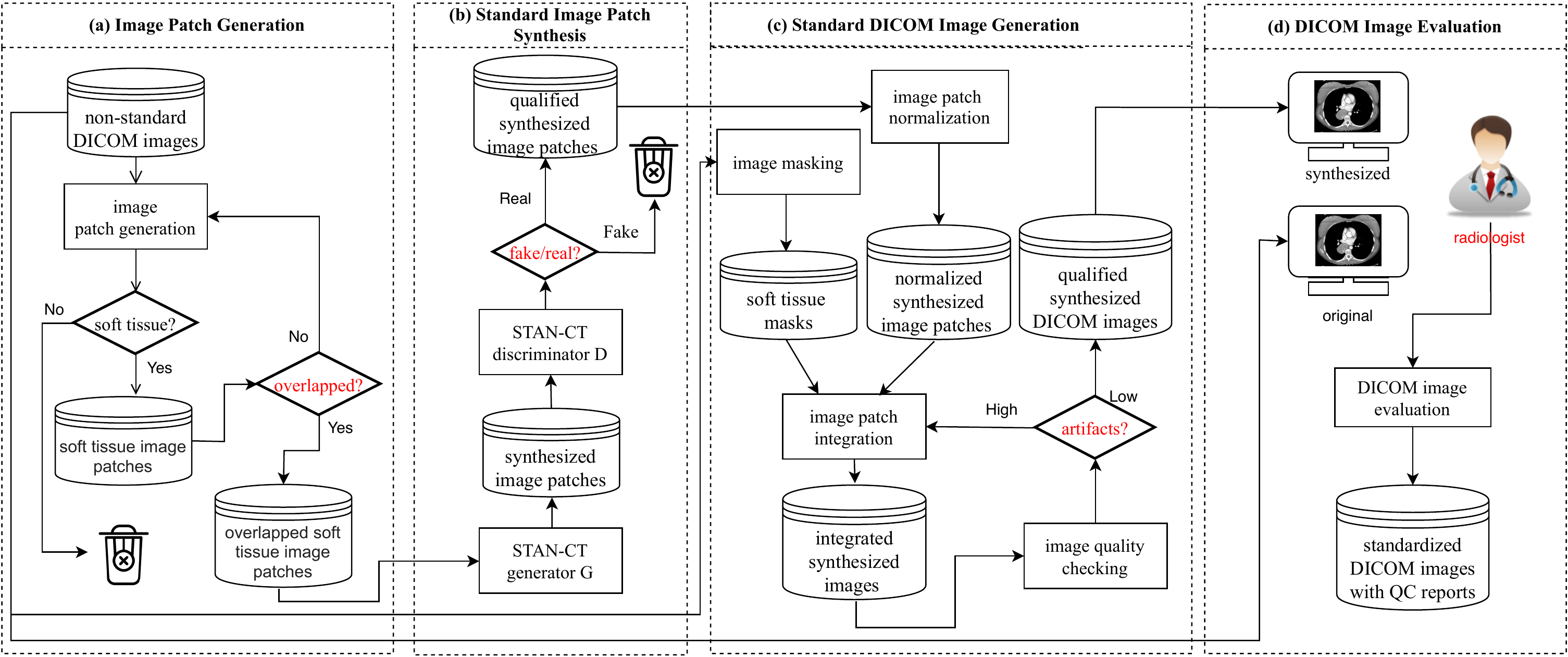}
\caption{{\bf STAN-CT DICOM-to-DICOM image standardization framework.} (a) Soft tissue image patches are generated from the input DICOM files, ready for image standardization and normalization. (b) For all the soft tissue image patches, new image patches are synthesized using STAN-CT generator. The quality of the new image patches is checked using STAN-CT discriminator. (c) All the synthesized soft-tissue image patches are integrated and are filtered by a soft tissue image mask generated using the input DICOM image. DICOM image quality is checked by examining box artifacts and empty pixels. (d) The synthesized and the original non-standard DICOM image files will be viewed side-by-side by radiologists using a PACS reading workstation. The radiologists’ reports will be used to further evaluate the quality of the standardized CT images. Meanwhile, image texture features will be extracted for automatic performance evaluation. } \label{fig:framework}
\end{figure}

\subsection*{DICOM Reconstruction Framework}
STAN-CT presents a DICOM-to-DICOM reconstruction framework for systematic DICOM reconstruction. While the core of the framework is the GAN model introduced in Section~\ref{sec:GAN}, the DICOM-to-DICOM reconstruction framework includes four additional components to facilitate processes such as image patch generation and fusion (see Fig.~\ref{fig:framework}). Note that each component has a unique quality control unit (red diamond box) that ensures the outputs are free from defects.

\noindent {\bf Step 1. soft tissue image patch generation:}
The first step of STAN-CT DICOM-to-DICOM image standardization is soft tissue image patch generation. Image patches with size between 100 and 256 are randomly generated using the input DICOM image. An image patch is a soft tissue image patch if at least 70\% of the pixels are in the soft tissue range (Hounsfield unit value ranging from -1000 to 900). The process will continue until each soft-tissue image patch contains at least 50\% overlapped pixels.

\noindent {\bf Step 2. standard image patch synthesis:}
With a trained STAN-CT generator, a soft-tissue image patch obtained in the previous step will be standardized (see Section~\ref{sec:GAN} for details). 
Then, the synthesized image patches will be examined by STAN-CT discriminator. If a synthesized image patch can fool the discriminator, it is considered as a \textit{qualified synthesized image patch}. Otherwise, the synthesized image patch will be discarded. This step ensures the quality of the synthesized image patches. 

\noindent {\bf Step 3. standard DICOM image generation:}
Given all the qualified synthesized image patches, we first normalize the pixel intensity from gray-scale to the Hounsfield unit using: 
\begin{equation}
P_{HU} = \frac{P_g - min(\hat{x}_g)}{max(\hat{x}_g)-min(\hat{x}_g)} . (MAX - MIN)
\end{equation}
\noindent where $P_{HU}$ and $P_{g}$ is the pixel value in Hounsfield unit and gray-scale unit respectively, $\hat{x}_g$ is a qualified synthesized image patch, and $MAX$ and $MIN$ are the maximum and minimum CT number of a source DICOM.

Meanwhile, with a soft tissue image mask created from the original DICOM images with Hounsfield unit ranging from $-1000$ to $900$, the non-soft tissue parts of the synthesized and normalized image patches will be discarded. 
Finally, we integrate all the valid soft tissue patches to generate the integrated synthesized images. 

The quality of the integrated synthesized images will be checked using a quality control unit, which inspects whether there is any box artifacts or  missing values. If some artifacts are identified, the corresponding image patches will be re-integrated by cropping boundary pixels.

\noindent {\bf Step 4. DICOM image evaluation:}
Here, we evaluate DICOM image quality manually and automatically. First, both the synthesized and the original non-standard DICOM image files will be viewed side-by-side by radiologists using a PACS reading workstation. Radiologists will be asked to evaluate image quality, estimate the acquisition protocol, and extract tumor properties. The radiologists’ reports will be used to manually evaluate the quality of the standardized CT images. 
Meanwhile, with all the synthesized DICOM files generated in the previous step, image texture features will be automatically extracted and compared for  performance evaluation.

\section{Experimental result}
\subsection*{Data}
For the training data, we used total of 14,688 CT image slices captured using three different kernels (BL57, BL64, and BR40) and four different slice thicknesses (0.5, 1, 1.5, 3mm) using  Siemens CT Somatom Force scanner at the University of Kentucky Medical Center. STAN-CT adopted BL64 kernel and 1mm slice thickness as the standard protocol since it has been widely used in clinical practice. 
Random cropping was used for the image patch extraction and resized into $256\times256$ pixel patches. Data augmentation was done by rotating and shifting image patches. Finally, a total of 49,000 soft-tissue image patches were generated from the CT slices and were used as the training data of STAN-CT.

Two testing data sets were prepared for STAN-CT performance evaluation. Both data sets were captured using Siemens CT Somatom Force scanner at the University of Kentucky Medical Center hospital. The first testing data were captured using the non-standard protocol BR40 and 1mm slice thickness. The second testing data were captured using the non-standard protocol BL57 and 1mm slice thickness. The image patch generation step was the same as that of the training data. 
Each test data set contains 3,810 image patches.  

\subsection*{STAN-CT architecture and hyperparameters}
STAN-CT GAN model consists of a U-net with fifteen hidden layers and an FCN with five hidden layers. The $4\times4$ kernel is used in the convolutional layer. LeakyRelu~\cite{xu2015empirical} is adopted as the activation function in all the hidden layers. Softmax is used in the last layer of FCN. Random weight is used during the network initialization phase. The prediction thresholds for determining fake or real images is 0.01 and 0.99 respectively. Maximum training epochs were set to 100 with a learning rate of 0.0001 with momentum 0.5. A fully optimized alternative training mechanism (the same as GANai) was used for the network training.
STAN-CT was implemented in TensorFlow~\cite{tensorflow2015-whitepaper} on a Linux computer server with eight Nvidia GTX 1080 GPU cards. The model took about 36 hours to train from scratch. Once the model was trained, it took about 0.1 seconds to synthesize and normalize every image patch.

\subsection*{Evaluation Metric}
For performance evaluation, we computed five radiomic texture features (i.e. dissimilarity, contrast, homogeneity, energy, and correlation) using Gray Level Co-occurrence Matrix (GLCM). 
The absolute error $\mathbb{E}$ of each radiomic texture feature was computed using:

\begin{equation}
\mathbb{E}{(I_{syn}, I_{target}, f)}  = \frac{| \varphi{(I_{syn}, f)}- \varphi{(I_{target}, f)} |}{\varphi{(I_{target}, f)}}
\end{equation}
\noindent where $\varphi$ is the GLCM feature extractor $I_{syn} and I_{target}$ is the synthesized image from STAN-CT and the target image respectively. $f$ is the corresponding feature space.

\subsection*{Performance of image patch synthesis}
\begin{figure}
\centering
\includegraphics[width=\textwidth]{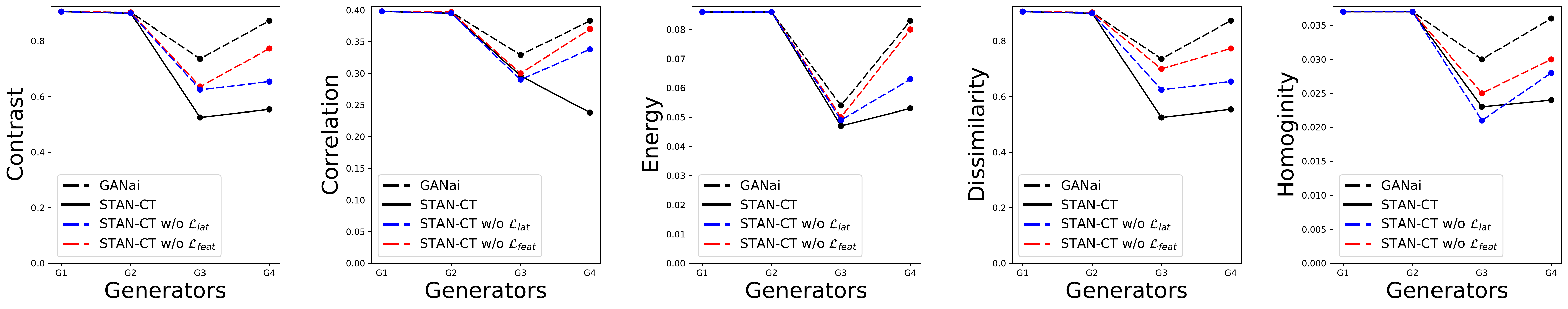}
\caption{{\bf Performance evaluation of $G$.} Generators produced in the first four iterations of alternative training of STAN-CT or GANai were compared using the GLCM-based features. The X-axis denotes the version of the training phase, and the Y-axis denotes the absolute error of each selected texture feature. The result indicates STAN-CT archived overall the best performance. } 
\label{fig:g_comps}
\end{figure}

Table~\ref{tab:fetures} shows the absolute error of five GLCM-based texture features of STAN-CT, GANai (the current state-of-the-art model), and two disentangled representation of STAN-CT. In the model named ``STAN-CT  w/o $\mathcal{L}_{lat}$'', we discarded from STAN-CT the latent loss function $\mathcal{L}_{lat}$ of $G$. In the second one named ``STAN-CT w/o $\mathcal{L}_{feat}$'', we discarded the feature-based loss $\mathcal{L}_{feat}$ of $D$ from STAN-CT. 
All the models were tested using kernel BL57 and BR40 with the same slice thickness ($1 mm$). For kernel BL57, STAN-CT and its variants outperformed GANai in all the texture features. For kernel BR40, STAN-CT was significantly better than GANai in four out of five  features.  
The first four generators of each GAN models were selected for further analysis. 
Fig.~\ref{fig:g_comps} illustrates the change of the absolute errors of the five GLCM-based texture features using the generators produced in the first four iterations of alternative training of STAN-CT or GANai. The result indicates that STAN-CT can quickly reduce the errors in the first a few iteration of the alternative training, while no clear trend was observed in the results of GANai.

%


\subsection*{Performance of DICOM reconstruction}


\begin{table}
\small
\centering
\makebox[0pt][c]{\parbox{1\textwidth}{%
    \begin{minipage}[b]{0.45\hsize}\centering
        \begin{tabular}{llllll}
            \hline
            Kernel	&	Features	&	\rotatebox[origin=c]{90}{GANai}	&	\rotatebox[origin=c]{90}{STAN-CT}	&	\rotatebox[origin=c]{90}{\makecell{STAN-CT  \\  w/o  $\mathcal{L}_{Lat}$}}	&	\rotatebox[origin=c]{90}{\makecell{STAN-CT  \\ w/o $\mathcal{L}_{feat}$}}		\\ \hline
            \multirow{5}{*}{BL57}	&	dissimilarity	&	0.313	&	\textbf{0.228}	&	0.234	&	0.245	\\
            	&	contrast	&	0.313	&	\textbf{0.228}	&	0.234	&	0.245	\\
            	&	homogeneity	&	0.012	&	\textbf{0.009}	&	0.009	&	0.012	\\
            	&	energy	&	0.032	&	0.035	&	0.038	&	\textbf{0.022}	\\
            	&	correlation	&	0.085	&	0.058	&	\textbf{0.057}	&	0.120	\\ \hline
            											
            \multirow{5}{*}{BR40}	&	dissimilarity	&	0.683	&	\textbf{0.407}	&	0.441	&	0.545	\\
            	&	contrast	&	0.683	&	\textbf{0.407}	&	0.441	&	0.545	\\
            	&	homogeneity	&	0.028	&	\textbf{0.018}	&	0.019	&	0.024	\\
            	&	energy	&	\textbf{0.040}	&	0.041	&	0.057	&	0.045	\\
            	&	correlation	&	0.315	&	\textbf{0.203}	&	0.273	&	0.303	\\ \hline
            
            \hline
            \end{tabular}
        \caption {{\bf Texture feature comparison between GANai, STAN-CT and its two variants.} Five texture features (dissimilarity, contrast, homogeneity, energy and correlation) were extracted from DICOM image patches. The absolute error was reported for each feature.} 
        \label{tab:fetures} 
    \end{minipage}
    \hfill
    \begin{minipage}[b]{0.48\hsize}
            \begin{tabular}{llllll}
            \hline
            Kernel	&	Features	&	\rotatebox[origin=c]{90}{\makecell{ straight-\\forward}}	&	\rotatebox[origin=c]{90}{\makecell{ w/ overlapp-\\ed  check }}	&	\rotatebox[origin=c]{90}{\makecell{w/ real/fake \\ check} }	&	\rotatebox[origin=c]{90}{\makecell{STAN-CT \\ }}	\\  \hline
            \multirow{5}{*}{BL57}	&	dissimilarity	&	0.727	&	0.485	&	0.334	&	\textbf{0.201}	\\
            	&	contrast	&	0.727	&	0.485	&	0.334	&	\textbf{0.201}	\\
            	&	homogeneity	&	0.031	&	0.019	&	0.012	&	\textbf{0.009}	\\
            	&	energy	&	0.072	&	0.063	&	0.046	&	\textbf{0.032}	\\
            	&	correlation	&	0.319	&	0.149	&	0.075	&	\textbf{0.054}	\\ \hline
            \multirow{5}{*}{BR40}	&	dissimilarity	&	0.849	&	0.653	&	0.496	&\textbf{0.405}	\\
            	&	contrast	&	0.849	&	0.653	&	0.496	&	\textbf{0.405}	\\
            	&	homogeneity	&	0.035	&	0.027	&	0.022	&\textbf{0.016}	\\
            	&	energy	&	0.048	&	0.045	&	0.051	&	\textbf{0.040}	\\
            	&	correlation	&	0.386	&	0.345	&	0.289	&	\textbf{0.201}	\\ 
            \hline
            \end{tabular}
        \caption {{\bf Texture feature comparison.} Five texture features were extracted from DICOM images constructed from the same image patches using four different DICOM reconstruction methods. The averaged absolute error is reported for each feature.} 
        \label{tab:dicom} 
    \end{minipage}
    
}}
\end{table}

A straightforward patch-based image reconstruction approach has three steps: 1) splitting a DICOM slice into overlapped or non-overlapped image patches; 2) standardizing each image patch; and 3) merging the standardized image patches into one DICOM slice. A common problem in such a patch-based image reconstruction process is image artifacts, such as boundary artifact or inconsistent texture. As shown in Table~\ref{tab:dicom}, the straightforward approach has the highest absolute error on all the tested image features. 

\begin{figure}
    \centering
    \begin{subfigure}{0.5\textwidth}
        \centering
        \includegraphics[height=1.2in]{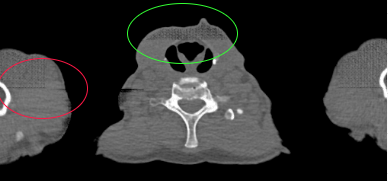}
        \caption{DICOM reconstructed using a straightforward method. The red circle highlights the boundary effect where two image patches were merged. The green circle shows texture inconsistency.}\label{fig:g_dicom}
    \end{subfigure}%
    ~ 
    \begin{subfigure}{0.5\textwidth}
        \centering
        \includegraphics[height=1.2in]{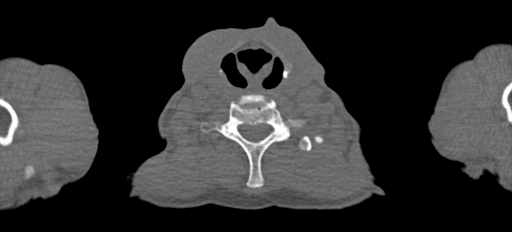}
        \caption{The same DICOM reconstructed using STAN-CT. No visual artifacts were found according to the radiologist's report.} \label{fig:s_dicom}
    \end{subfigure}
    \caption{\bf DICOM Reconstruction comparison} \label{fig:dicom}
\end{figure}

In STAN-CT, three quality control units were inserted into the framework, each being adopted to address a specific image quality problem. Table~\ref{tab:dicom} shows that STAN-CT achieved significantly better performance than the straightforward method regarding the absolute errors on five selected texture features. Fig.~\ref{fig:dicom} visualized the reconstructed DICOM images using the two methods. The red (green) circle highlights the boundary effect where two image patches were merged (texture inconsistency within a DICOM slice) using the straightforward method. In the same DICOM reconstructed using STAN-CT, no visual artifacts were found according to the radiologist's report.

Also, we compared STAN-CT with its two variants. The method named ``w/ overlapped  check'' used only the first quality control unit to check whether there were enough overlapped soft-tissue image patches. The method named ``w/ real/fake check'' used the first two quality control units, which not only checked if there were enough image patches, but also examined whether the image patches were successfully standardized. Table~\ref{tab:dicom} shows that both approaches achieve better results than the straightforward method, but none of is better than STAN-CT, indicating all the three quality control units are critical regarding artifact detection and removal. 
The standardized DICOM images, along with the corresponding standard images, were reviewed by radiologists at the Department of radiology, University of Kentucky using the picture archiving and communication system (PACS) viewer (Barco, GA, USA).  The radiologists, who were blinded to the image reconstruction algorithms, reported that no obvious difference was observed in lung regions between the two kinds of images.

\section{Discussion}
By systematically removing every single component in the GAN model and in the DICOM reconstruction pipeline using the the leave-one-out approach, we analyzed the impact of every component of STAN-CT. 

\begin{figure}
\centering
\includegraphics[width=.6\textwidth]{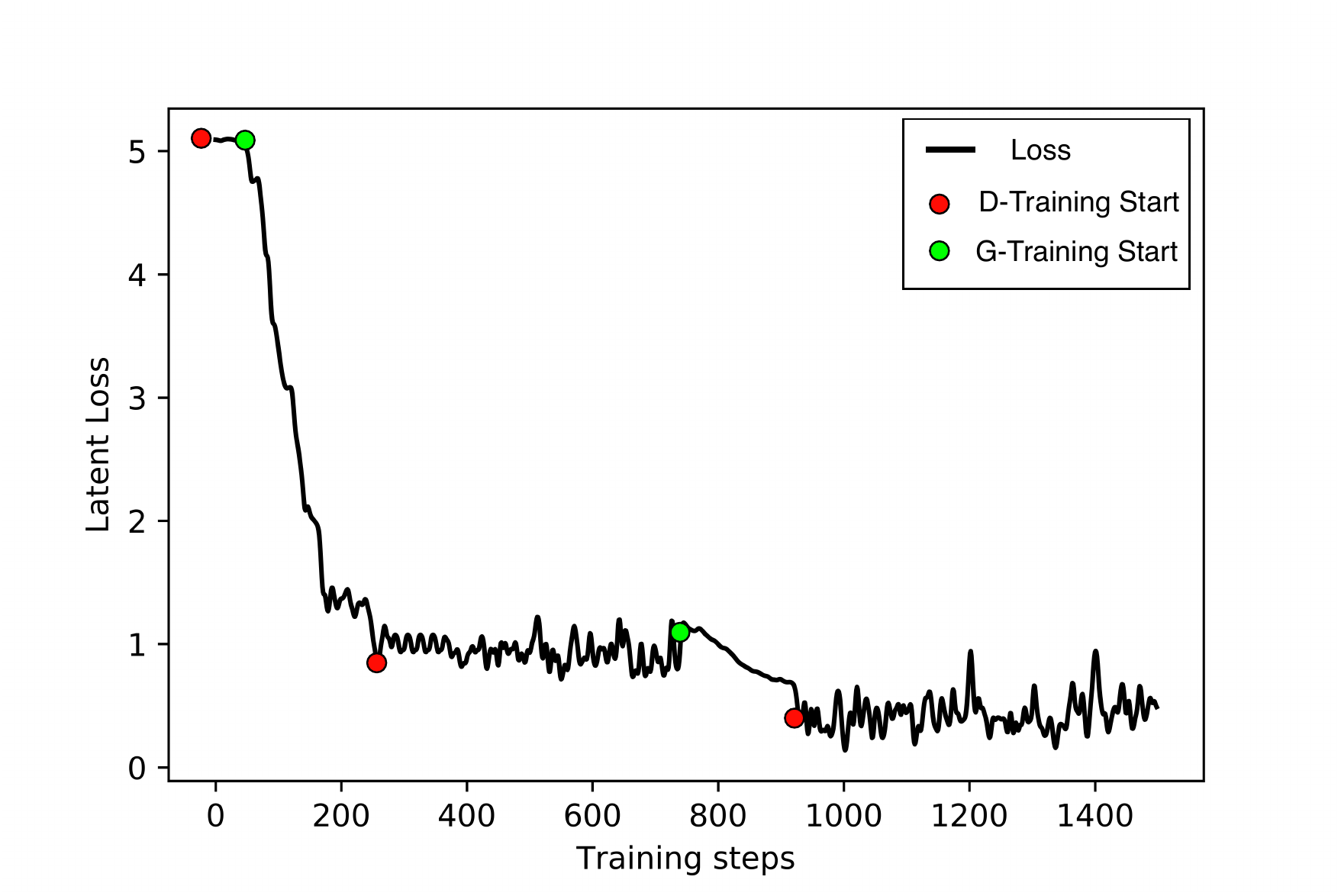}
\caption{{\bf Latent loss during GAN training.} Latent loss is decreasing effectively in the G-training phase, while it keeps stable in the D-training phase. } 
\label{fig:lat_loss}
\end{figure}

In STAN-CT, both the latent loss $\mathcal{L}_{Lat}$ and the feature loss $\mathcal{L}_{feat}$ are key components. To evaluate the impact of the loss functions, two versions of STAN-CT GAN, where the latent loss or the feature loss has been removed respectively, were created. Table~\ref{tab:fetures} shows that none of them can achieve the same performance as that of STAN-CT regarding the GLCM-based texture features. 
In addition, Figure~\ref{fig:lat_loss} shows that the latent loss  of STAN-CT $\mathcal{L}_{Lat}$ decreases during G-training, indicating that the generator can reduce the gap between the distributions of the target image and the synthetic image effectively, while maintaining flat during the D-training phases. 

The DICOM reconstruction pipeline includes four quality control units, each contributing to the improvements of the quality of the resulting DICOM images. Table~\ref{tab:dicom} shows that the contrast error of the straightforward DICOM reconstruction (without using any of the quality control units) is 0.727, which can be reduced to 0.485 by adding the overlapped soft tissue quality control, which provides consistent texture throughout the DICOM. It can be further reduced to 0.334 (54\% improvement) by adding the discriminator checker that ensures the success of image synthesis. Eventually, if all the four quality control units were used, the contract error was reduced to 0.201 (72\% improvement). 
In summary, our experiments demonstrate that STAN-CT is a robust framework for CT image standardization. 

\section{Conclusion}
Data discrepancy in CT images due to the use of non-standard image acquisition protocols adds extra burden to radiologists and also creates a gap in large-scale cross-center radiomic studies. 
We propose STAN-CT, a novel tool for CT DICOM image standardization and normalization. In STAN-CT, new loss functions are introduced for efficient GAN training, and a dedicated DICOM-to-DICOM image reconstruction framework has been developed to automate the DICOM standardization and normalization process. The experimental results show that STAN-CT is significantly better than the existing tools on CT image standardization.

\section*{Acknowledgements}
This research is supported by NIH NCI (grant no. 1R21CA231911) and Kentucky Lung Cancer Research (grant no. KLCR-3048113817).

\makeatletter
\small
\renewcommand{\@biblabel}[1]{\hfill #1.}
\makeatother

\bibliographystyle{unsrt}
\bibliography{main}

\end{document}